\documentclass[nobm,debug]{epl}
\usepackage{graphics}

\newcommand{\eg} {{\it e.g., }}

\newcommand{\half} {\frac{1}{2}}
\newcommand{\ie} {{\it i.e., }}
\newcommand{\rmd} {{\rm d}}
\newcommand{\rme} {{\rm e}}

\title{Unstable topography of biphasic surfactant monolayers}

\author{H. Diamant\inst{1}\thanks{E-mail:
\email{diamant@control.uchicago.edu}} \and T. A. Witten\inst{1,2}
\and A. Gopal\inst{3} \and K. Y. C. Lee\inst{3}}

\shortauthor{H. Diamant \etal}

\institute{
  \inst{1} The James Franck Institute \\
  \inst{2} Department of Physics \\
  \inst{3} Department of Chemistry and Institute for Biophysical
  Dynamics \\
  The University of Chicago, Chicago, IL 60637, USA
}


\begin{document}

\maketitle

\begin{abstract}

We study the conformation of a heterogeneous
surfactant monolayer at a fluid-fluid interface, near a boundary
between two lateral regions of differing elastic properties. The
monolayer attains a conformation of shallow, steep `mesas' with a
height difference of up to 10 nm. If the monolayer is progressively
compressed (\eg in a Langmuir trough), the profile develops 
overhangs and finally becomes unstable at a surface tension of 
about $K(\delta c_0)^2$, where $\delta c_0$ is the difference in
spontaneous curvature and $K$ a bending stiffness. We discuss the
relevance of this instability to recently observed folding
behavior in lung surfactant monolayers, and to the absence of
domain structures in films separating oil and water in emulsions.

\end{abstract}

Insoluble (Langmuir) monolayers of amphiphilic molecules, lying
at water-air or water-oil interfaces, have been extensively
studied in the past decades \cite{review_mono}.
Such monolayers are found in many applications, including, \eg
surface-tension reduction, emulsification and coating.
Of particular interest are phospholipid monolayers, which
are used in various studies to model the surface of cell
membranes \cite{review_lipid}.
Lipid monolayers are encountered in other biological
systems, such as the lung surfactant monolayer coating the
alveoli in lungs \cite{review_lung}.

An interesting issue is the departure of a monolayer, upon lateral
compression, from a flat, two-dimensional conformation to a
buckled, three-dimensional one. This aspect is particularly
important in the case of lung surfactant monolayers, which undergo
compression/expansion cycles during breathing. The buckling
transition was theoretically studied in previous works 
\cite{MJP,Saint,Granek}.
These studies focused on the overall conformation of the entire
monolayer, seeking an instability with respect to a single
extended mode (or a prescribed combination of modes \cite{Granek})
of undulation. They yield a buckling transition at a practically
vanishing surface tension (\ie very high compression).

In many circumstances the monolayer is inhomogeneous. The
heterogeneity may arise in single-component monolayers from the
coexistence of expanded and condensed domains \cite{domains}; 
multi-component monolayers may phase-separate to form domains of 
different composition.
The coupling
between conformation and inhomogeneous composition was thoroughly
studied as well
\cite{Granek,Andelman1,Andelman2,Wang,Kodama,Fred93,Lipowsky},
usually focusing on bilayer membranes. These works considered
annealed variations in composition, leading to spontaneous
formation of ripples, modulated phases and shape transformations.

Unlike symmetric bilayers, lipid monolayers usually have
a finite spontaneous curvature, arising either from
the asymmetry of the lipid molecule itself, or from electrostatic
interactions (\ie the dielectric asymmetry between the aqueous
and the non-polar air or oil phases) \cite{Glen}.
The common picture is that below the buckling
transition the monolayer conformation is flat despite the
spontaneous curvature.
The reason is that the bending energy per unit area
to be gained by a curved conformation
(typically a few tenths of $k_{\rm B}T$ per nm$^2$,
$k_{\rm B}T$ being the thermal energy),
is much smaller than the required work against surface 
tension (usually more than 10 $k_{\rm B}T$ per nm$^2$).
Hence, only at very low tension (\ie high compression)
is the monolayer expected to depart from a flat conformation
and buckle.
This argument, however, applies to the overall
spatial behavior of a homogeneous monolayer.

In the current work we ask a different question: what is the {\em
local} response of a Langmuir monolayer to a fixed
profile in mechanical properties 
as arising from a lateral domain structure.
There are four important length scales in the system:
the typical domain size, $L$, the thickness of a domain boundary, 
$d$, the spontaneous radius of curvature, $c_0^{-1}$, and the 
elastic length, $\lambda=(K/\gamma)^{1/2}$ determining the lateral 
length scale of height variations ($K$ being the bending rigidity and 
$\gamma$ the surface tension).
The discussion in this Letter is much simplified by assuming
$L\gg\lambda\gg d$ (the value of $c_0$ remains unrestricted).
Since $L$ is of order 10 $\mu$m and $\lambda$ is typically 1--10
nm, the first inequality is well justified. It allows us to 
isolate a single, straight boundary, separating two infinite 
domains.
The second assumption, $\lambda\gg d$, allows us to 
treat the domain boundary as infinitely sharp. 
In practice, for a biphasic layer far from its critical point,
$d$ is typically the size of a few headgroups, roughly 1 nm 
(though it may  exceed the scale of $\lambda$ in some circumstances).  
For typical $d$ we find that the infinitely sharp limit is a good 
approximation \cite{ournext}.
An additional requirement is that the monolayer surface 
maintain its integrity throughout the compression. This is generally 
not the case in practice; lipid monolayers often fracture and form
multi-layers at a finite pressure \cite{multilayers}. This occurs
as the monolayer yields to vertical shear stresses, brought about
by the lateral pressure combined with small curvature. Another
mechanism encountered in practice is the ejection of vesicles into
the aqueous phase \cite{ejection}. However, the presence of
certain additives in natural and model lung surfactants was found
to suppress these microscopic types of collapse \cite{lung}.

Using these observations, we can find the shape
of the surface as a function of its projected area or surface tension,
without the usual assumption of a moderate, single-valued
height function.
Thus, we implicitly take into account all modes of response and 
allow for overhangs.
(A related calculation was previously presented for studying
different domain shapes \cite{Harden}. Being restricted to linear
response, this model yielded mild height modulations in the form
of stable `caplets', rather than the sharp `mesas' and
conformation instability found in the current work.)

\begin{figure}
\centerline{\resizebox{0.45\textwidth}{!}
{\includegraphics{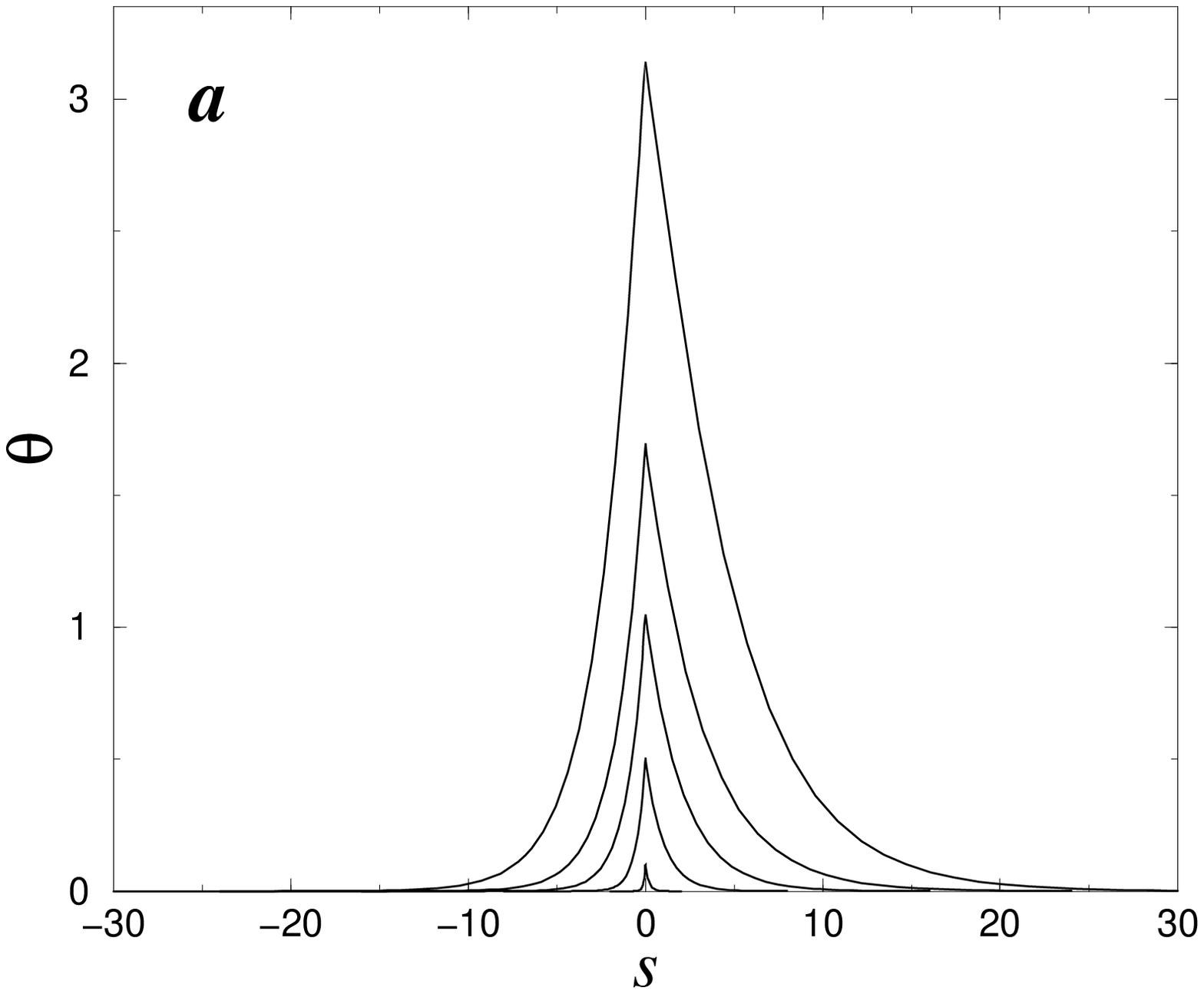}} 
\hspace{.2cm}
\resizebox{0.463\textwidth}{!}
{\includegraphics{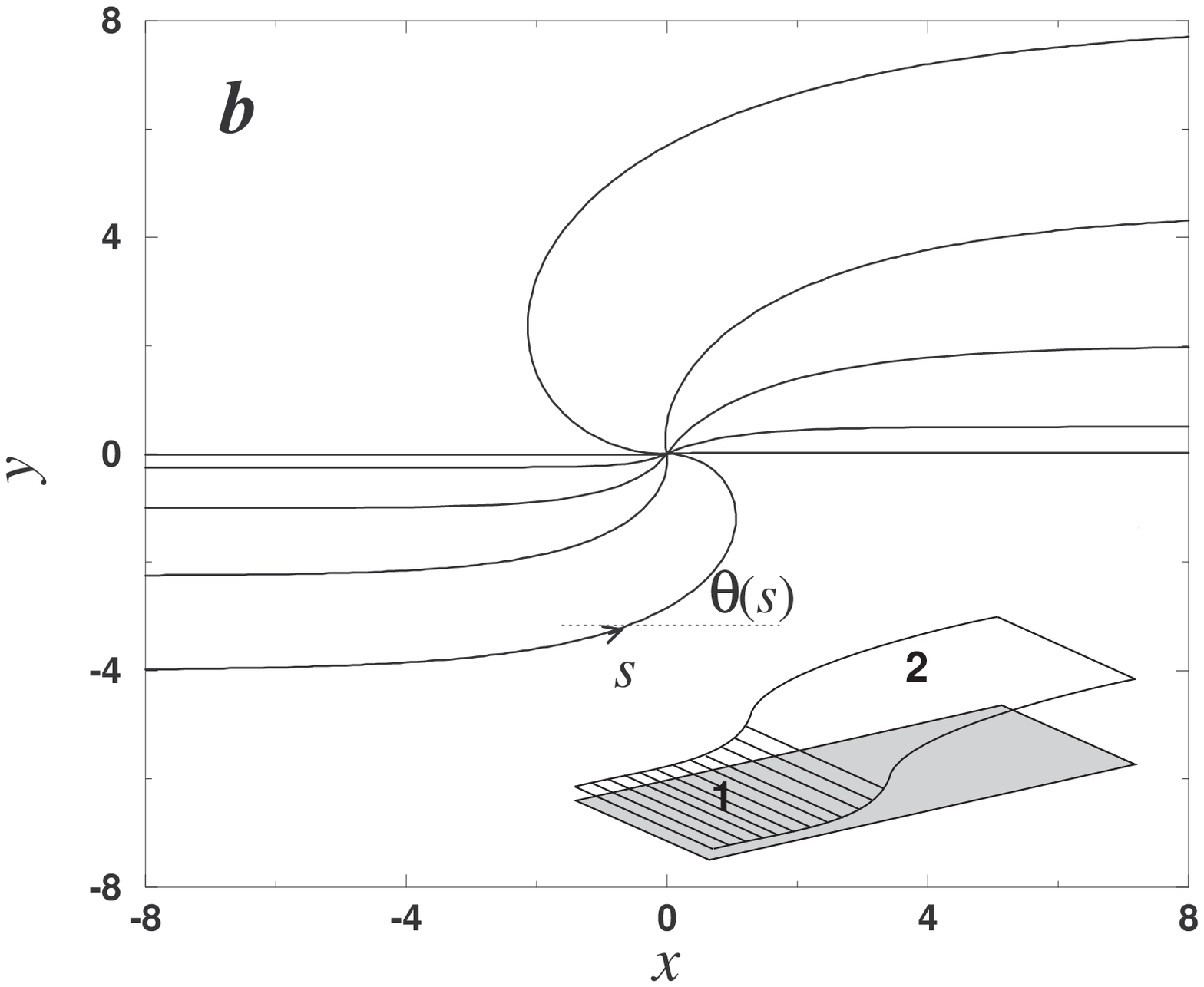}}} 
\vspace{-0.3cm}
\caption{{\it a}) Slope angle profiles
near a domain wall as compression is increased. The curves are
obtained from eqs.\ (\ref{profile})--(\ref{theta0}) 
using the following parameters
(from bottom to top): $v=0.1, 0.5, 1, 1.5, 2$; $\lambda_1=0.1,
0.5, 1, 1.5, 2$; $\lambda_2=0.2, 1, 2, 3, 4$. (The proportions
$v$:$\lambda_1$:$\lambda_2$ are kept fixed so as to simulate a
compression process, decreasing $\gamma$ while keeping all other
parameters constant.) {\it b}) The corresponding spatial 
conformations.
The units of length are arbitrary (they are typically of order 10
nm). The inset shows a schematic three-dimensional sketch of 
the monolayer shape for $\theta_0<\pi/2$.} 
\label{fig_compress}
\end{figure}

We represent the boundary region of two large domains as a surface 
whose left and right halves differ in bending rigidity $K$ and spontaneous 
curvature $c_0$.  On the left half $K=K_1$ and $c_0=c_{01}$; on the 
right half the values are $K_2$ and $c_{02}$.  
The resulting surface is 
uniform in the direction parallel to the domain boundary but curved in the 
perpendicular direction, as shown in fig.~\ref{fig_compress}b.  
We may thus represent the surface by its intersection with a vertical 
plane perpendicular to the boundary.  
We define the monolayer conformation by the local angle $\theta(s)$ 
between the surface and a reference plane
at curvilinear distance $s$ from the domain boundary 
(see fig.~\ref{fig_compress}b).
The bending energy $U$ of the monolayer can thus be written as
$
  U = L \int_{-\infty}^0 \rmd s [\half K_1 \dot\theta^2(s) - 
  K_1 c_{01}\dot\theta(s)] +
  L \int_0^\infty \rmd s [\half K_2 \dot\theta^2(s) - 
  K_2 c_{02}\dot\theta(s)]
$.
%
Here $L$ is the length of the domain boundary and $\dot\theta \equiv
\rmd\theta/\rmd s$ is the local curvature. 
We require that $\theta(s)$ be continuous at $s=0$, so that the 
surface is smooth everywhere.  
The state of minimum $U$ (without including tension) is a curled surface 
with curvature $c_{01}$ on the left and $c_{02}$ on the right. 
Only a net tension $\gamma$ in the monolayer allows the surface to 
approach a flat conformation. 
This tension adds a term $-\gamma A_{\rm p}$ to the energy, 
where $A_{\rm p}$ is the projected area of the surface, shown 
as a shaded plane in fig.~\ref{fig_compress}b. 
The element of projected area is $\rmd s \cos\theta$, so that the 
full energy to be minimized can be written as
$
  G \equiv U + \gamma L \int_{-\infty}^\infty \rmd s 
  [1 - \cos\theta(s)]
$.
%

Thus our system is equivalent to a pre-curled sheet of paper of length 
$L$ joined at one edge to a more strongly curled sheet and then subjected 
to a tensile force $\gamma L$ acting so as to straighten the curling.  
Even under large tensions, there is a nonzero departure from the flat 
state.
To demonstrate this departure we consider the bending moments of the two 
sheets across the junction line. 
The left side exerts a moment on the right side equal to
$\delta U/\delta\dot\theta(0^-) = LK_1[\dot\theta(0^-) - c_{01}]$.  
This must equal the moment acting from the right side, 
$LK_2[\dot\theta(0^+) - c_{02}]$.
Hence the curvatures on the two sides are in general unequal: 
$K_1\dot\theta(0^-) - K_2\dot\theta(0^+) = K_1 c_{01} - K_2 c_{02} 
\equiv \Delta$, where the parameter $\Delta$ (having dimensions of 
force) characterizes the extent of heterogeneity.
The nonzero curvature $\dot\theta(0)$ at the boundary relaxes to zero 
on a distance $\lambda$ (to be determined), producing a net slope angle 
$\theta_0\equiv\theta(s=0)\sim\lambda\dot\theta(0)$ with bending energy
$KL\lambda\dot\theta(0)^2$.
For small $\theta_0$ the associated loss of projected area is of order
$L\lambda\theta_0^2$, with energy 
$\gamma L\lambda [\lambda\dot\theta(0)]^2$. 
The decay length $\lambda$ is that which minimizes the total energy, 
\ie $\lambda\sim(K/\gamma)^{1/2}$ as anticipated above.
For $K_1=K_2$ the angle $\theta_0$ is proportional to the difference 
in spontaneous curvature, $\delta c_0\equiv c_{01}-c_{02}$:
$\theta_0 \sim (\Delta/K)\lambda \sim (K/\gamma)^{1/2}\delta c_0$.

To show the explicit profile and the buckling instability, 
we rewrite the energy per unit length by integrating the linear 
term in $\dot\theta$ while requiring that the monolayer be flat
far away from the boundary (\ie at the centers of the two
contiguous domains), $\theta(s\rightarrow\pm\infty)=0$,
\begin{equation}
  g[\theta(s)] \equiv G/L = \int\rmd s [(K/2) \dot{\theta}^2
  + \gamma(1-\cos\theta)] - \theta_0\Delta.
\label{SineGordonH}
\end{equation}
Equation (\ref{SineGordonH}) has the familiar form of
a physical pendulum action (with $s$ as imaginary time). 
Variation with respect to
$\theta(s\neq 0)$ gives the Sine-Gordon equation,
%
  $K_i\ddot{\theta} = \gamma\sin\theta$, $i=1,2$, 
%
whose first integration yields
%
  $\dot{\theta}^2 = 4\lambda_i^{-2}\sin^2(\theta/2)$.
%
Second integration leads to
soliton profiles on both sides of the boundary,
\begin{equation}
  \tan(\theta/4) = \left\{ \begin{array}{ll}
  \tan(\theta_0/4) \rme^{s/\lambda_1} \ \ \ \ & s<0 \\
  \tan(\theta_0/4) \rme^{-s/\lambda_2} & s>0
  \end{array}\right.
\label{profile}
\end{equation}
Finally, the condition for the jump in curvature at the
boundary (\eg as found above from a moment balance argument, 
or by variation of $g$ with respect to $\theta_0$) is
$K_1\dot{\theta}(0^-)-K_2\dot{\theta}(0^+)=\Delta$,
which determines $\theta_0$ as
\begin{equation}
  \sin(\theta_0/2) = 
  \Delta/[2\sqrt{\gamma}(\sqrt{K_1}+\sqrt{K_2})]
  \equiv v/2.
\label{theta0}
\end{equation}
Thus, for any finite $\Delta$, the monolayer has
a sigmoidal shape whose maximum slope angle is given
by $\theta_0$ of eq.~(\ref{theta0}). 
The total height difference is given by
\begin{equation}
  h = \int_{-\infty}^\infty\rmd s\sin\theta =
  \Delta/\gamma.
\label{height}
\end{equation}
Substituting the obtained profile back in eq.~(\ref{SineGordonH}),
we find the energy of the conformation,
\begin{equation}
  {g}/{\Delta} = 2\tan(\theta_0/4) - \theta_0
  = 4(1-\sqrt{1-v^2/4})/v - 2\sin^{-1}(v/2).
\label{conf_energy}
\end{equation}
The relation between the projected area and surface tension
is found by imposing the area constraint,
$A_{\rm p}=L\int\rmd s\cos\theta$, or, equivalently, by taking
the derivative of $g$ with respect to $\gamma$,
\begin{equation}
  \delta L \equiv (A-A_{\rm p})/L =
  \partial g/\partial\gamma
  = [2(\sqrt{K_1}+\sqrt{K_2})^2/\Delta] v(1-\sqrt{1-v^2/4}).
\label{dL}
\end{equation}

According to eqs. (\ref{theta0})--(\ref{dL}), as compression is
increased, \ie $\delta L$ is increased or $\gamma$ is decreased
(depending on the experimental scenario), the step profile becomes
sharper (larger $\theta_0$), higher (larger $h$), and more
favorable (lower $g$). The process is demonstrated in
fig.~\ref{fig_compress}.
At a certain stage $\theta_0$ becomes larger than $\pi/2$ and an
overhang forms. However, as is evident from
eqs.~(\ref{theta0})--(\ref{dL}), there is a critical value of
compression beyond which the profile equations have no solution.
Beyond this point our physical pendulum goes over the top and
our model surface curls up.
This happens when $\theta_0$ reaches $\pi$, corresponding to
$v_{\rm c}=2$, or
\begin{equation}
  \gamma_{\rm c} = [\Delta/2(\sqrt{K_1}+\sqrt{K_2})]^2
  \simeq (K/16)(\delta c_0)^2,\ \ \ \ 
  \delta L_{\rm c} = 4(\sqrt{K_1}+\sqrt{K_2})^2/\Delta
  \simeq 16/|\delta c_0|.
\end{equation}
(The approximate expressions assume that the heterogeneity is
mainly manifested in different $c_0$ rather than different $K$.)
The dimensions of the step are finite at the critical compression,
$h_{\rm c}=\delta L_{\rm c}\simeq 16/\delta c_0$. Yet, the lateral
compressibility diverges,
%
  $\partial\delta L/\partial\gamma \sim (v_{\rm c}-v)^{-1/2}$
as $v\rightarrow v_c$,
%
implying instability; extra area can be pulled into the inflected
region without resistance, and the monolayer tries to curl up.
(Detailed description of this critical response, however, is
beyond the scope of the current model.)
Macroscopically, the instability should show up as a plateau
in the pressure--area isotherm of the monolayer.

Substituting typical values for phospholipid monolayers 
\cite{Safran} ---
$\gamma\simeq$ 10--50 erg/cm$^2$, $K\simeq$ 10--50 $k_{\rm B}T$,
$c_0^{-1}\simeq$ 5--10 nm --- we get $\lambda\simeq$ 1--10 nm, 
$v\simeq$ 0.1--1, and $h\simeq$ 0.1--10 nm. 
%
Hence, the mesas are sharp but
shallow. The numerical value of $v\sim 1$ implies that the
predicted instability may be encountered for 
attainable pressures. 
Furthermore, the
energy per unit length gained by departing from a flat
conformation is $g\simeq\Delta\simeq$ 1--10 $k_{\rm B}T$/nm
(or a few piconewtons). 
Hence, for a typical domain size of about 1--10 $\mu$m, 
the sigmoidal conformation is robust to thermal 
fluctuations.\footnote{
One might worry about the gravitational energy cost
of displacing water from the flat interface.
This energy per unit area is about
$\delta\rho gh^2 \sim 10^4$ $k_{\rm B}T/\mbox{cm}^2$,
where $\delta\rho$ is the difference in density of the two
phases and $g$ the gravitational acceleration.
Thus, due to the small height differences, gravity is
negligible over all relevant lateral length scales
(up to meters).
Beyond the instability, however, the monolayer may become
much more folded, and gravity may have a significant stabilizing
role.}

\begin{figure}
\centerline{\resizebox{0.45\textwidth}{!}
{\includegraphics{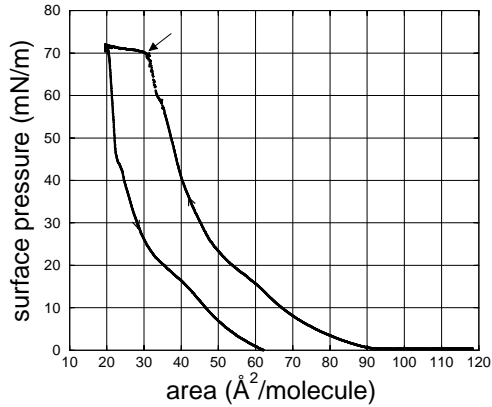}}} 
\vspace{-0.4cm}
\caption{Pressure--area isotherm for a mixed monolayer of 
DPPC and POPG, as measured during a compression/expansion cycle
in a Langmuir trough.
The mole ratio is DPPC:POPG$=$7:3 and the temperature 25$^\circ$C. 
The folding instability is indicated by an arrow.}
\label{fig_iso}
\end{figure}

\begin{figure}
\centerline{ {\normalsize\bf{\it a}} \resizebox{0.3\textwidth}{!}
{\includegraphics{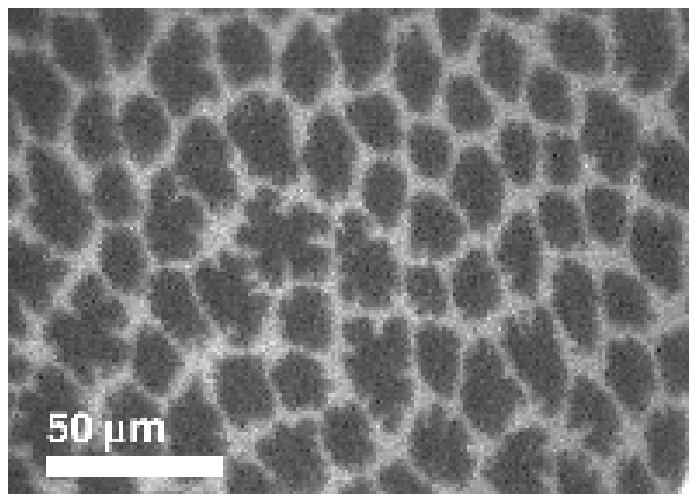}}
\hspace{.1cm} {\normalsize\bf{\it b}} \resizebox{0.3\textwidth}{!}
{\includegraphics{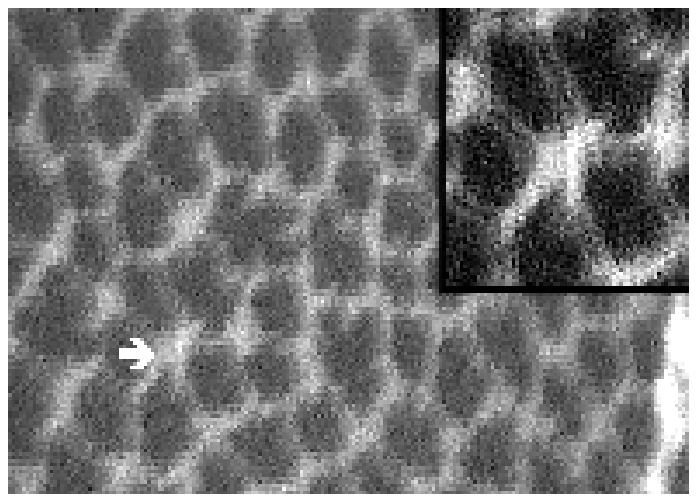}}}
\vspace{.2cm}
\centerline{ {\normalsize\bf{\it c}} \resizebox{0.3\textwidth}{!}
{\includegraphics{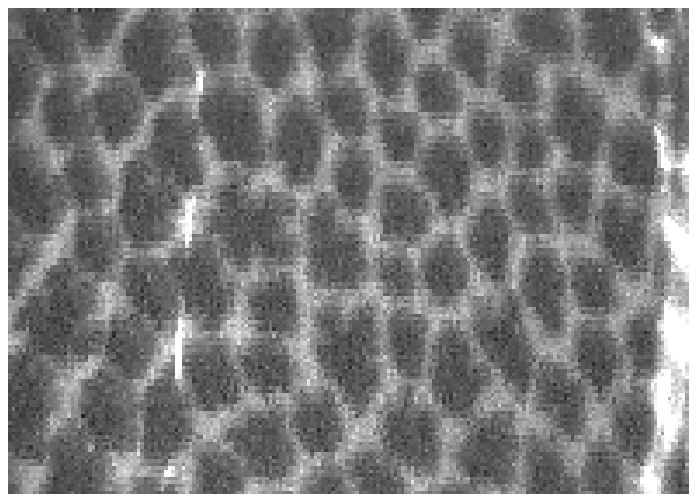}}
\hspace{.1cm} {\normalsize\bf{\it d}} \resizebox{0.3\textwidth}{!}
{\includegraphics{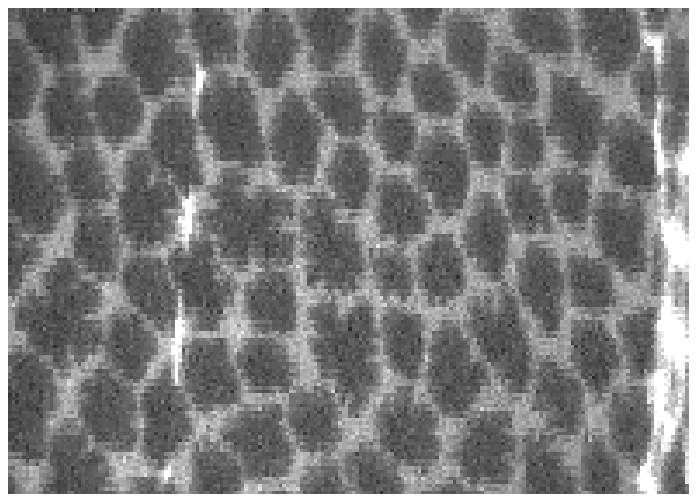}}}
\caption{Fluorescence microscopy images of the folding instability.
{\it a}) Section of the monolayer just before folding ($t=0$), 
exhibiting the biphasic domain structure. Dark regions are DPPC-rich;
bright ones are POPG-rich.
{\it b}) The same section at $t=$1/30 sec. A micron-scale fold
appears in between domain walls (indicated by arrow). 
The image is blurred because
of monolayer movement during folding. The inset shows a 
contrast-enhanced image of the fold,
magnified by 50 percent.
{\it c}) The fold at $t=$2/30 sec, having propagated to nearby
domains.
{\it d}) The fold at $t=$4/30 sec, after the fast monolayer 
movement has ceased.}
\label{fig_fold}
\end{figure}

In recent experiments on model lung surfactant monolayers a new
type of instability has been observed \cite{lung}. As the
monolayer is compressed into the coexistence region, containing
domains of different composition, there is a critical lateral
pressure at which it locally folds towards the aqueous phase.
Similar folding has been observed in simpler phospholipid mixtures
as well \cite{lung}.
Figure~\ref{fig_iso} shows a pressure--area isotherm as measured
for a mixed monolayer of 
dipalmitoylphosphatidylcholine (DPPC) and 
palmitoyloleoylphosphatidylglycerol (POPG). 
The folding
is manifested by a plateau in the isotherm, occurring, for this
system, at a very low surface tension. (The same phenomenon,
however, was observed in DPPG monolayers at a much higher surface
tension \cite{lung}.) Figure~\ref{fig_fold} presents a sequence of
fluorescence microscopy snapshots of the monolayer just before and
just after the instability. 
A micron-scale fold appears in between domain
walls, subsequently propagating to nearby domains.
The folding, as compared to other collapse mechanisms,
significantly reduces irreversibility and loss of surfactant
during a compression/expansion cycle.
It is believed, therefore, to be of importance for the function 
of lungs. 
We suggest that the observed folding 
might be initiated by the conformation instability
as found from the model. 
According to the model the folding should follow the
domain boundary, unlike the straight fold in fig.~\ref{fig_fold}.
Nevertheless, the surface shear viscosity of such monolayers is of 
order $1$ surface poise (dyn sec/cm) \cite{visco}.
For surface pressures smaller than $10^2$ dyn/cm this suggests
a shear relaxation time larger than $10^{-2}$ sec.
Since the fold forms on a time scale of about $10^{-2}$ sec 
(see fig.~\ref{fig_fold}), the monolayer may respond to the 
instability like an elastic sheet, thus inhibiting curved folds.
Certainly, more experiments are required before
a clear relation between the observed folding and theoretical
instability can be established.

In summary, we find that a Langmuir monolayer
should exhibit inflected profiles in the vicinity of
domain boundaries.
These profiles are the elastic response
to the contrast in spontaneous curvature and/or bending
rigidity between the two domains.
For a monolayer composed of many coexisting domains, this
leads to an overall conformation of `mesas',
where the domains of higher $Kc_0$ stick down towards the
aqueous phase.
As the monolayer is compressed, the mesas grow more
pronounced, subsequently developing small overhangs at their edges,
and finally becoming unstable.
Within the current model we could obtain the folding point
only from an instability criterion. In practice, the folded
structure may become favorable before the instability 
(\ie for $\theta_0<\pi$), whereupon folded regions would 
coexist with unfolded ones.
(In addition, the shape instability might be pre-empted 
by other collapse mechanisms, as mentioned in the introduction.)

Apart from the folding instability,
two rather general conclusions arise from this analysis.
First, a biphasic Langmuir monolayer at a water-air or
water-oil interface should practically never be completely flat; a
conformation of mesas should appear for any finite surface
tension. 
Such mesas are expected to exist, therefore, in many 
experimental and natural systems.
Conversely, monolayers at water-oil interfaces in
emulsions, microemulsions or L$_3$ phases \cite{Safran}, whose
surface tension practically vanishes, cannot have contiguous
regions of differing elastic properties. 
At a vanishing tension any such heterogeneity
would necessarily yield a shape instability. Indeed, to the best
of our knowledge, no microdomain structures have ever been
observed in those systems, as opposed to Langmuir monolayers.

Several issues remain to be examined.
One is the stability of a
straight domain boundary (\ie possible rippling of the mesa wall).
Another is the effect of introducing a domain boundary of
finite thickness.
These issues will be addressed in a forthcoming publication
\cite{ournext}.
Finally, it should be noted that the theoretical findings
presented here do not conform with the
conventional picture of {\em flat}
lipid monolayers below the buckling transition.
Since the model relies on very few, plausible assumptions,
we believe that the inferred conformation of mesas
should be observed in practice.
Such an observation, however, may be difficult in view of
the small height differences involved, and the fluid interface on
which the monolayer is deposited.
We hope that this work will motivate experiments in
this new, intriguing direction.

\acknowledgments 
We benefited from discussions and correspondence
with D. Andelman, J. Klein and S. Safran.
This work was supported by the National Science Foundation
under Awards Nos.\ DMR 9975533 and 9728858, and by its 
MRSEC program under Award No.\ 9808595.
HD was partially supported by the American Lung Association
(RG-085-N).
AG was supported by the Searle Scholars Program/The Chicago
Community Trust (99-C-105).
The experimental apparatus was made possible by an NSF 
CRIF/Junior Faculty Grant (CHE-9816513).
KYCL is grateful for support from the March of Dimes Basil
O'Connor Starter Scholar Research Award (5-FY98-0728), and
the David and Lucile Packard Foundation (99-1465).


\end{document}